\documentstyle[twoside,fleqn,espcrc2,epsf]{article}
\newcommand{\be}{\begin{equation}}
\newcommand{\ee}{\end{equation}}
\newcommand{\ba}{\begin{eqnarray}}
\newcommand{\ea}{\end{eqnarray}}
\def\reff#1{(\ref{#1})}
\newcommand{\Tr}{\mathop{\rm Tr}\nolimits}

\newcommand{\1}{1\!\!\!\bot}
\def\vsig{\vec{\mbox{$\sigma$}}}
\def\vca{\vec{\mbox{$\cal A$}}}
\title{Discretization effects and gauge independence
 for the electric and magnetic screening masses\thanks{Talk
 presented by A.\ Cucchieri. This work has been supported
 by the TMR network ERBFMRX-CT-970122 and the DFG under
 grant Ka 1198/4-1.}}

\author{Attilio Cucchieri and Frithjof Karsch\address{Fakult\"at
        f\"ur Physik, Universit\"at Bielefeld, D-33615 Bielefeld,
        GERMANY}}
       
\begin{document}

\begin{abstract}
We evaluate the electric and magnetic screening masses from the
long-distance behavior of the (temporal and spatial) gluon 
correlation functions, for pure $SU(2)$ gauge theory at finite
temperature. In order to investigate the gauge dependence of the
screening masses we consider seven different gauges. We also 
evaluate these masses using different definitions of the lattice
gluon field, corresponding to discretization errors of different
orders.
\end{abstract}

\maketitle

%%%%%%%%%%%%%%%%%%%%%%%%%%%%%%%%%%%%%%%%%%%%%%%%%%%%%%%%%%%%%%%%%%%%%%%

\section{INTRODUCTION}

The electric and magnetic screening masses can be obtained from the 
exponential decay of finite-temperature gluon correlation functions 
at large separations \cite{masses,masses2}. It is important to
stress that, even though the gluon propagator is a gauge dependent
quantity, the (pole) masses obtained from it are --- within a wide 
class of gauges --- gauge invariant to arbitrary order in perturbation
theory \cite{Kobes}.

Here we present a preliminary study of the gluon propagator at 
finite temperature, using different definitions of the lattice 
gluon field, and considering several different gauges, in order to
check for discretization effects and gauge independence \cite{masses2}
of the screening masses.

%%%%%%%%%%%%%%%%%%%%%%%%%%%%%%%%%%%%%%%%%%%%%%%%%%%%%%%%%%%%%%%%%%%%%%%%

\section{LATTICE SETUP}
\label{sec:setup}

We consider a standard Wilson action for $SU(2)$ lattice gauge theory
in $4$ dimensions, with periodic boundary conditions. (Details of
notation and numerical simulations will be given in \cite{future}.)
We evaluate the lattice gluon propagators 
\ba
D_{\mu \mu}(z) & \!\!\! \equiv & \!\!\! 
\frac{1}{3\,V\,N_T}
\sum_{x_3\mbox{,}\, b}\,\langle\,
Q_{\mu}^{b}(x_3 + z)\, Q_{\mu}^{b}(x_3)\,\rangle\;\mbox{.}
\nonumber
\ea
Here $V \equiv N_1 N_2 N_3$,
\ba
Q_{\mu}^{b}(x_3)
& \!\!\! \equiv & \!\!\! 
\sum_{x_0\mbox{,}\,x_{\perp}}
\, A_{\mu}^{b}(x_0\mbox{,}\,x_{\perp}\mbox{,}\,x_3)
\;\mbox{,}
\nonumber
\ea
$ A_{\mu}^{b}(x) \equiv \Tr \left[ \, A_{\mu}(x) \,
\sigma^{b} \, \right] / 2 $, and $\sigma^{b}$ is a Pauli matrix. The
lattice gauge field $A_{\mu}(x)$ will be defined in Section 
\ref{sec:discre} below.

From the exponential decay at large separation of the (temporal and
spatial) gluon correlation functions we can obtain the electric and
magnetic screening masses \cite{masses,masses2}:
\ba
D_{00}(z) & \!\!\! \sim & \!\!\! \exp\left( - M_e z \right) \nonumber \\
D_{ii}(z) & \!\!\! \equiv & \!\!\! D_{11}(z) + D_{22}(z) 
 \sim \exp\left( - M_m z \right) \;\mbox{.} \nonumber
\ea
As mentioned before, these screening masses are expected to be
gauge invariant \cite{Kobes}.

%%%%%%%%%%%%%%%%%%%%%%%%%%%%%%%%%%%%%%%%%%%%%%%%%%%%%%%%%%%%%%%%%%%%%%%%

\subsection{Gauge Fixing}
\protect\vspace{0.2cm}

We consider the ``so-called'' $\lambda$-gauge condition \cite{lambda},
i.e.\ we look for a local minimum of the functional
\be
{\cal E}_{U}^{\lambda}[ g ] 
        \equiv  -\sum_{x\mbox{,}\,\mu} 
   \Tr\Bigl[\lambda_{\mu} g(x)U_{\mu}(x)g^{\dagger}(x + e_{\mu})
\Bigr]
\,\mbox{,}
\label{eq:min}
\ee
\noindent
where $\lambda_0 = \lambda$ and $\lambda_i = 1$ for $i = 1\mbox{,}\,2
\mbox{,}\,3$.
We have performed simulations with $\lambda = 2\mbox{,} \, 1\mbox{,}\,
0.5\mbox{,}\, 0.1\mbox{,}\, 0.01$ and $\,0$. The case $\lambda=1$ is
the Landau gauge, whereas $\lambda=0$ is the Coulomb gauge. In the last 
case we also impose the condition
\ba
Q_{0}^{b}(x_0) & \!\!\! \equiv & \!\!\! \sum_{x_{\perp}\mbox{,}\,x_3}
\, A_{0}^{b}(x_0\mbox{,}\,x_{\perp}\mbox{,}\,x_3)
\, = {\cal Q}_{0}^{b}
\;\mbox{,}
\nonumber
\ea
which is automatically satisfyed for any $\lambda \neq 0$.

Finally, we consider the standard Maximally Abelian gauge. In this case
we fix the residual $U(1)$-gauge symmetry by imposing a $U(1)$-Landau 
gauge condition \cite{KS}.

%%%%%%%%%%%%%%%%%%%%%%%%%%%%%%%%%%%%%%%%%%%%%%%%%%%%%%%%%%%%%%%%%%%%%%%%

\subsection{Discretization Effects}
\label{sec:discre}
\protect\vspace{0.2cm}

The gauge field is usually defined as
\ba
A_{\mu}^{(1)}(x) &\!\!\! \equiv & \!\!\!
\left[\,U_{\mu}(x)\,-\,U_{\mu}^{\dagger}(x)\,\right]
\, / \, (2\,i)
\;\mbox{,}
\label{eq:Amux}
\nonumber
\ea
where $U_{\mu}(x) \in SU(2)$ are link variables. We can
also consider other definitions of $A_{\mu}(x)$, leading to 
discretization errors of different orders. For example, we can 
write \cite{Giusti,CM}
\ba
 A_{\mu}^{(2)}(x) &\!\!\! \equiv & \!\!\!
\left\{\,
 \left[ U_{\mu}(x) \right]^2\,-\,\left[ U_{\mu}^{\dagger}(x)
\right]^2\,\right\}
\, / \, (4\,i)
\nonumber \\
 A_{\mu}^{(3)}(x) &\!\!\! \equiv & \!\!\! \left\{ \,
 \left[ U_{\mu}(x) \right]^4\,-\,\left[ U_{\mu}^{\dagger}(x)
\right]^4\,\right\}
\, / \, (8\,i) \;\; \mbox{.} \nonumber
\ea
If we set $ \, U_{\mu}(x)\equiv \exp[i a g_0 \,\vsig \cdot \vca(x)]
\,\mbox{,}$ we obtain that $A_{\mu}^{(1)}(x)$, $A_{\mu}^{(2)}(x)$ and
$A_{\mu}^{(3)}(x)$ are equal to $\, a g_0 \,\vsig \cdot \vca(x) $ plus
terms of order $a^3\,g_0^3$. We can also consider \cite{CM}
\ba
A_{\mu}^{(4)}(x) & \!\!\!\! \equiv & \!\!\!\! \left[\,
  4 A_{\mu}^{(1)}(x) \,-\,
   A_{\mu}^{(2)}(x) \,\right]\,/\,3 \nonumber \\
A_{\mu}^{(5)}(x) & \!\!\!\! \equiv & \!\!\!\! \left[\,
  16 A_{\mu}^{(1)}(x) \,-\,
   A_{\mu}^{(3)}(x) \,\right]\,/\,15 \nonumber \\
A_{\mu}^{(6)}(x) & \!\!\!\! \equiv & \!\!\!\! \left[\,
  4 A_{\mu}^{(2)}(x) \,-\,
   A_{\mu}^{(3)}(x) \,\right]\,/\,3 \nonumber \\
A_{\mu}^{(7)}(x) & \!\!\!\!\!\! \equiv & \!\!\!\!\!\!
    \left[ 64 A_{\mu}^{(1)}(x) -
              20 A_{\mu}^{(2)}(x) +
                 A_{\mu}^{(3)}(x)\right] / 45 \nonumber \,\mbox{.}
\ea
It is easy to check that $A_{\mu}^{(4)}(x)$, $A_{\mu}^{(5)}(x)$ and
$A_{\mu}^{(6)}(x)$ are equal to $\, a g_0 \,\vsig \cdot \vca(x) $ plus
terms of order $a^5\,g_0^5$, and that $A_{\mu}^{(7)}(x) = a g_0 \,\vsig
\cdot \vca(x)$ plus terms of order $a^7\,g_0^7$. Finally,  by writing
$ U_{\mu}(x) \equiv U_{\mu}^{0}(x) \1 + i\, \vsig \cdot 
\vec{U}_{\mu}(x) $, one can define the ``perfect'' gluon 
field \cite{KS,Furui}
\ba
A_{\mu}^{(p) b}(x) & \!\!\! \equiv & \!\!\!
\frac{ U_{\mu}^{b}(x) }{\| \vec{U}_{\mu}(x) \|} \,
\arctan \left( \frac{\| \vec{U}_{\mu}(x) \|}{ U_{\mu}^{0}(x)} \right)
\;\mbox{,}
\nonumber
\ea
which is equal to $\, a g_0 \,\vsig \cdot \vca(x) $ with no 
discretization errors. 

Let us notice that the minimizing condition defined in
Eq.\ \reff{eq:min} implies
\ba
\sum_{\mu = 1}^{4} \,
\lambda_{\mu} \, \left[\,
{\cal A}_{\mu}^{b}(x) \,-\, {\cal A}_{\mu}^{b}(x - e_{\mu}) 
\,\right] & \!\!\! = & \!\!\!
 {\cal O}(a^2 g_0^2) \;
\mbox{.}
\nonumber
\ea

%%%%%%%%%%%%%%%%%%%%%%%%%%%%%%%%%%%%%%%%%%%%%%%%%%%%%%%%%%%%%%%%%%%%%

\section{SIMULATIONS AND RESULTS}

We have evaluated the gluon propagators $D^{(n)}_{ii}(z)$
and $D^{(n)}_{00}(z)$, using the eight different definitions of the 
gluon field $A_{\mu}^{(n)}(x)$ given in the previous section, 
for the seven gauges considered here, with two different sets of parameters:
({\bf s1}) $N_{T} = 4\mbox{,}\, V = 12^2$x$24 \mbox{,}\, \beta = 2.512$,
and ({\bf s2}) $N_{T} = 8\mbox{,}\, V = 16^2$x$32\mbox{,}\, \beta = 2.74$.
(In both cases $T \approx 2\,T_c$.) We note that we have generated
{\em different} sets of configurations for each different gauge, while, in
a given gauge, the different lattice discretizations of the gluon
propagators have been evaluated using the {\em same} set of configurations.

\begin{figure}[hbt]
\begin{center}
\protect\vspace*{-3.1cm}
\epsfxsize=0.42\textwidth
\protect\hspace*{0.4cm}
\leavevmode\epsffile{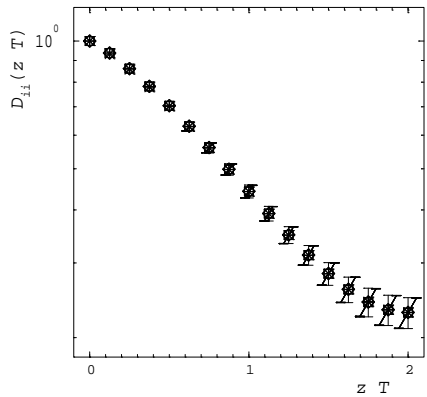}
\end{center}
\vspace*{-3.4cm}
\caption{~Data for the eight different discretizations of the
gluon propagator $D_{ii}(z T)$ in Landau gauge with simulation
parameters {\bf s2}. The propagators are normalized to 1 at $z T = 0$.
Error bars are one standard deviation.
}
\label{fig:discret}
\end{figure}

\setcounter{footnote}{0}
We have found that, in all gauges, the eight gluon correlation functions
$D^{(n)}(z)$ are equal modulo a constant factor, both in the electric
and in the magnetic sector.~\footnote{A similar result has also been 
obtained at zero temperature \cite{Giusti,CM}. Actually, for $\lambda = 
0.01$ and $0$, we have found that the propagator $D^{(3)}_{00}(z)$ is 
not proportional to the other seven propagators $D^{(n)}_{00}(z)$.
However, this discrepancy seems to decrease --- as expected --- if one
gets closer to the continuum limit, i.e.\ going from the {\bf s1} to 
the {\bf s2} simulation \protect\cite{future}.}
(In Fig.\ \ref{fig:discret} we plot, as an example, the data for the
magnetic case in Landau gauge.) This result
clearly implies that the screening masses are {\bf independent} of the
discretization of the gluon field.

\begin{figure}[hbt]
\begin{center}
\vspace*{-2.1cm}
\epsfxsize=0.42\textwidth
\protect\hspace*{0.4cm}
\leavevmode\epsffile{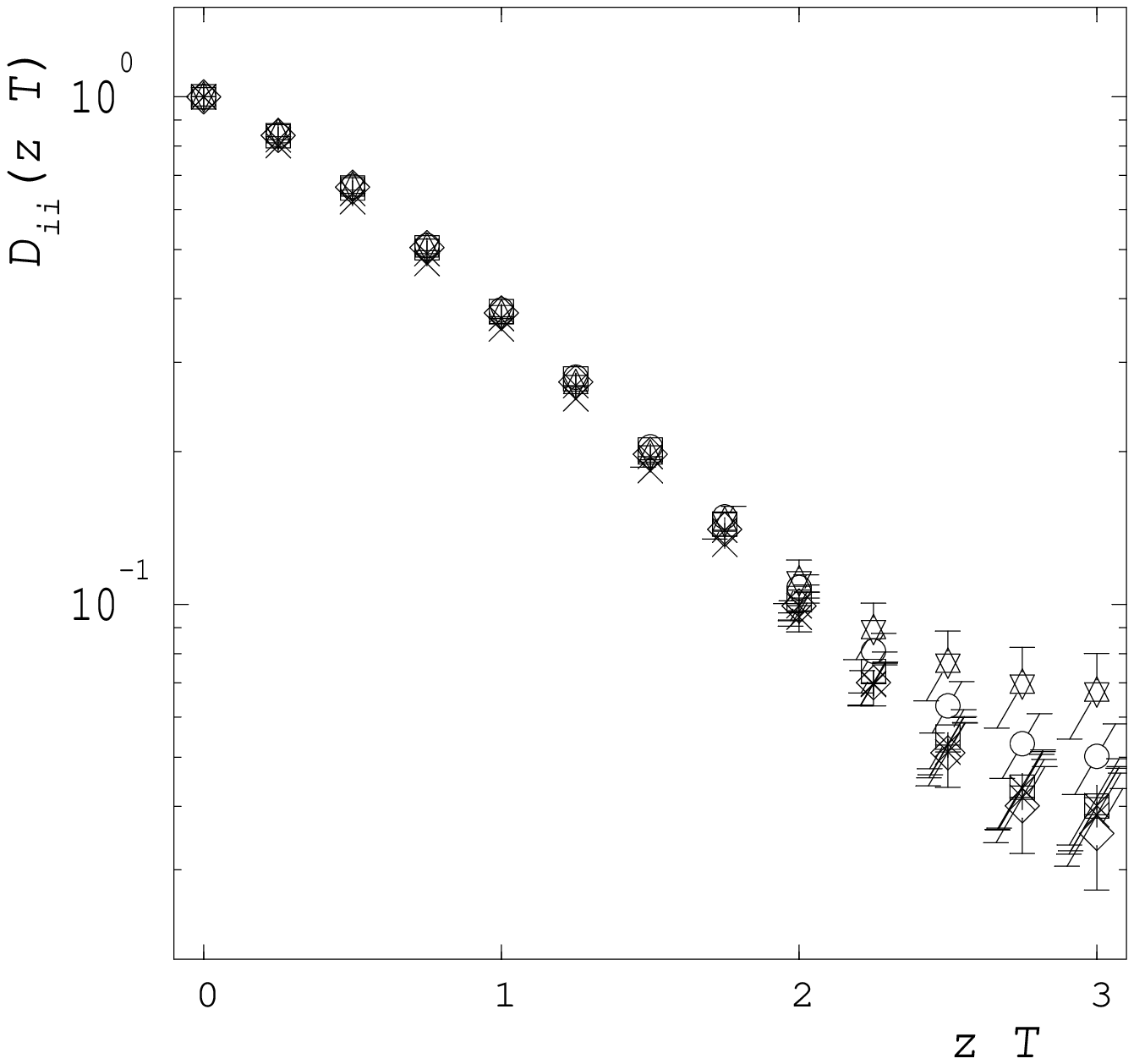}
\vspace*{-3.7cm}
\epsfxsize=0.42\textwidth
\protect\hspace*{0.4cm}
\leavevmode\epsffile{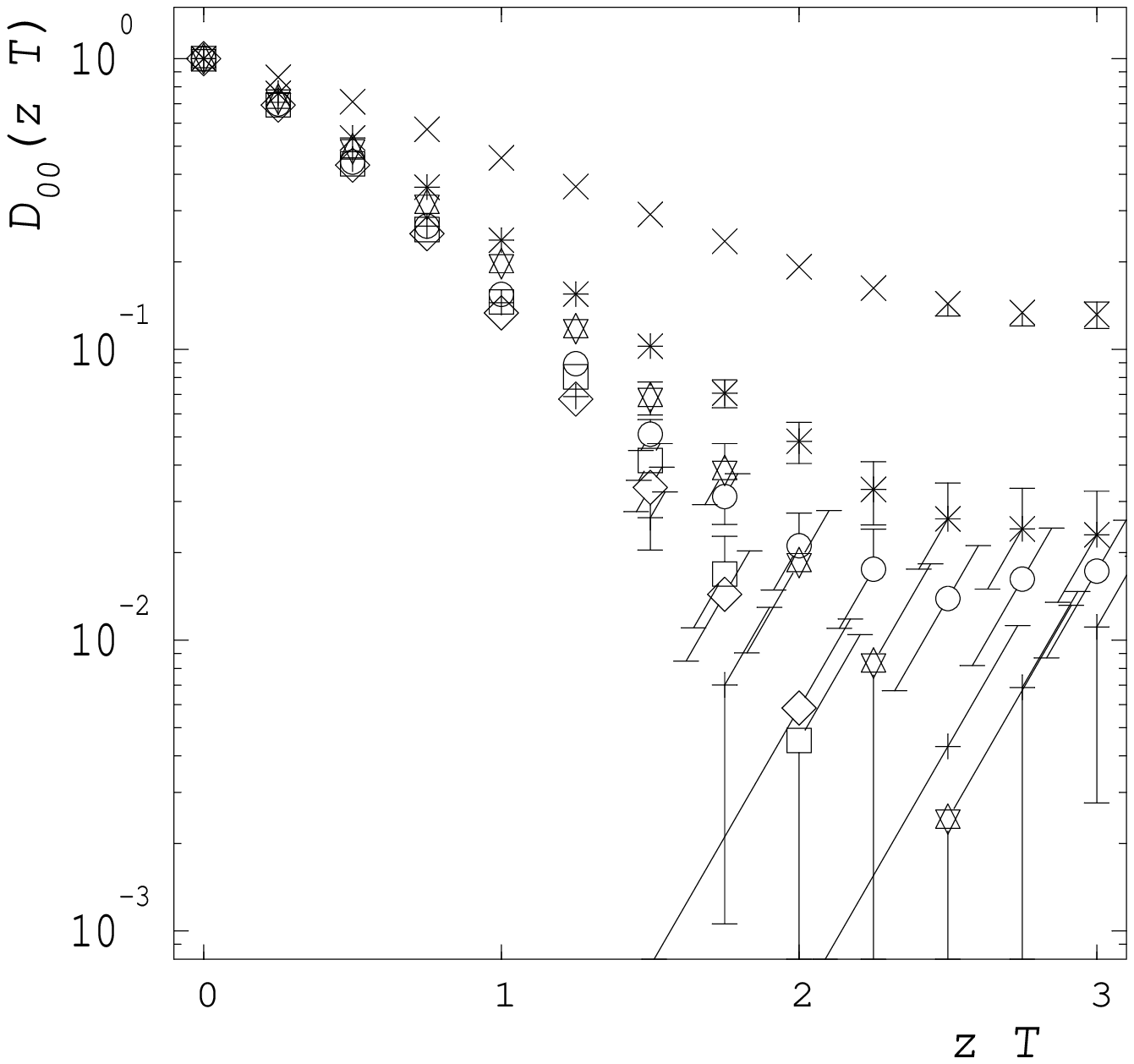}
\vspace*{-3.7cm}
\epsfxsize=0.42\textwidth
\protect\hspace*{0.4cm}
\leavevmode\epsffile{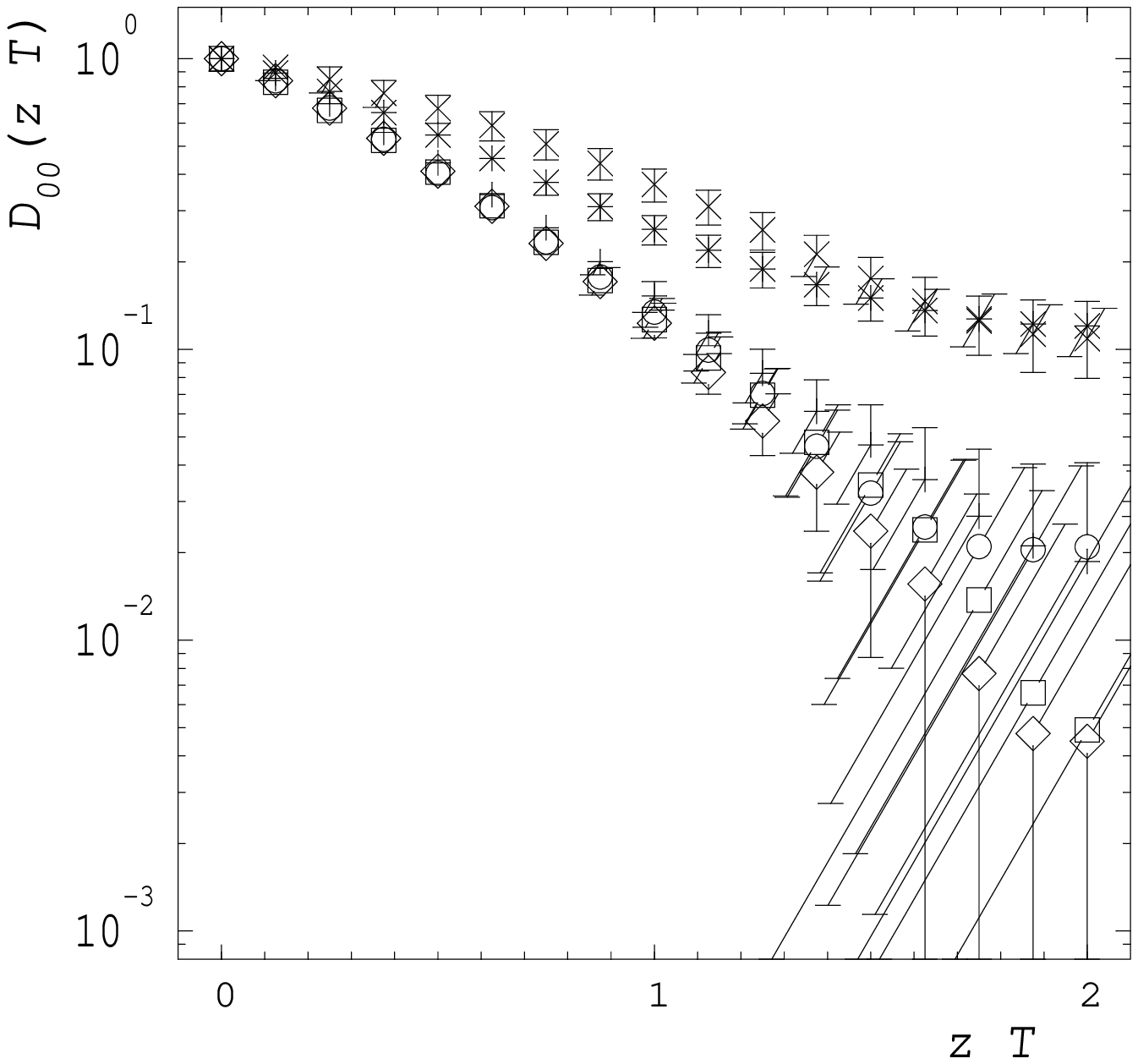}
\end{center}
\vspace*{-3.4cm}
\caption{~Data for the $D^{(1)}$ gluon propagator in the seven 
different gauges considered here:
$D_{ii}(z T)$ in Landau gauge with simulation parameters {\bf s1} 
(top figure); $D_{00}(z T)$ in Coulomb gauge with simulation
parameters {\bf s1} (center figure), and with simulation parameters 
{\bf s2} (bottom figure). The propagators are normalized to 1 at $z
T = 0$. Error bars are one standard deviation.
}
\label{fig:gauge}
\end{figure}

Since the different lattice discretizations of the gluon propagator
differ only by a constant factor, we can consider any of them
[for example $D^{(1)}(z)$] in order to compare the results 
obtained in different gauges. We then find that, in the magnetic
sector, the gluon correlation functions $D(z)$ corresponding
to different gauges are equal modulo a constant factor
(see Fig.\ \ref{fig:gauge}, top figure).
This clearly implies that the magnetic mass obtained in this
way is {\bf gauge independent}, 
at least within the seven gauges considered
here.
On the contrary, in the electric sector, the propagators 
corresponding to the gauges $\lambda = 0.01$ and $0$, are not
proportional to the propagators obtained in the other gauges
(see Fig.\ \ref{fig:gauge}, middle figure, 
symbols $\ast$ and $\times$). 
Moreover, the discrepancy between the Coulomb gauge propagator and the 
propagators in the $\lambda$-gauges does not decrease as one gets
closer to the continuum limit (see Fig.\ \ref{fig:gauge},
bottom figure). However, the strong variation seen in
the propagator for the case $\lambda=0.01$ (see Fig.\ \ref{fig:gauge},
middle and bottom figures) makes us expect that the
``gauge dependence'' seen for this small $\lambda$-value, as well as 
for $\lambda = 0$, may be related to our gauge-fixing procedure,
rather than indicating a gauge dependence of the 
electric mass. In fact, we observe that the gauge-fixing procedure 
does not always bring the $U_0 (x)$ fields close to unity, which 
is needed to ensure a meaningful definition of the $A_0(x)$ field
in terms of the compact gauge-field variables. We are currently 
investigating this issue in more detail \cite{future}.

%%%%%%%%%%%%%%%%%%%%%%%%%%%%%%%%%%%%%%%%%%%%%%%%%%%%%%%%%%%%%%%%%%%%%

%%%%%%%%%%%%%%%%%%%%%%%%%%%%%%%%%%%%%%%%%%%%%%%%%%%%%%%%%%%%%%%%%%%%%

%%%%%%%%%%%%%%%%%%%%%%%%%%%%%%%%%%%%%%%%%%%%%%%%%%%%%%%%%%%%%%%%%%%%%

\end{document}